\documentclass[10pt,letterpaper]{article}
\usepackage[top=0.85in,left=2.75in,footskip=0.75in]{geometry}
\usepackage{xr-hyper}

\usepackage{amsmath,amssymb}
\usepackage{changepage}
\usepackage{textcomp,marvosym}
\usepackage{cite}
\usepackage{nameref}
\usepackage{subcaption}
\usepackage[nopatch=eqnum]{microtype}
\DisableLigatures[f]{encoding = *, family = * }
\usepackage[table]{xcolor}
\usepackage{array}
\newcolumntype{+}{!{\vrule width 2pt}}
\newlength\savedwidth

\raggedright
\usepackage[aboveskip=1pt,labelfont=bf,labelsep=period,justification=raggedright,singlelinecheck=off]{caption}
\renewcommand{\figurename}{Fig}

\makeatletter
\renewcommand{\@biblabel}[1]{\quad#1.}
\makeatother

\usepackage{lastpage,fancyhdr,graphicx}
\usepackage{epstopdf}
\fancyheadoffset[L]{2.25in}
\fancyfootoffset[L]{2.25in}

\newcommand{\EP}[1]{\textcolor{red}{ {#1}}}
\renewcommand{\EP}[1]{ {#1}}

\usepackage{url}
\begin{document}

	\vspace*{0.2in}
	
	\begin{flushleft}
		{\Large
			\textbf\newline{Untrained Perceptual Loss for image denoising of line-like structures in MR images} 
		}
		\newline
		\\
		Elisabeth Pfaehler\textsuperscript{1*},
		Daniel Pflugfelder\textsuperscript{2},
		Hanno Scharr\textsuperscript{1},
		\\
		\bigskip
		\textbf{1} Institute for Advanced Simulation: Data Analytics and Machine Learning (IAS-8), Forschungszentrum J{\"u}lich GmbH, Germany, Wilhelm-Johnen Straße, J{\"u}lich, North-Rhine-Westfalia, Germany\\
		\textbf{2} Institute of Bio- and Geosciences: Plant Sciences (IBG-2), Forschungszentrum J{\"u}lich GmbH, Germany, Wilhelm-Johnen Straße, J{\"u}lich, North-Rhine-Westfalia, Germany
		\bigskip

		* e.pfaehler@fz-juelich.de
		
	\end{flushleft}
	\section*{Abstract}
	
	In the acquisition of Magnetic Resonance (MR) images shorter scan times lead to higher image noise. Therefore, automatic image denoising using deep learning methods is of high interest. MR images containing line-like structures such as roots or vessels yield special characteristics as they display connected structures and yield sparse information. For this kind of data, it is important to consider voxel neighborhoods when training a denoising network.  \\ 
	In this paper, we translate the Perceptual Loss to \EP{3D} data by comparing feature maps of untrained networks in the loss function as done previously for 2D data. We tested the performance of untrained Perceptual Loss (uPL) on \EP{3D} image denoising of MR images displaying brain vessels (MR angiograms - MRA) and images of plant roots in soil. We investigate the impact of various uPL characteristics such as weight initialization, network depth, kernel size, and pooling operations on the results. We tested the performance of the uPL loss on four Rician noise levels (1 \%, 5 \%, 10 \%, and 20 \%) using evaluation metrics such as the Structural Similarity Index Metric (SSIM).\\
	We observe, that our uPL outperforms conventional loss functions such as the L1 loss or a loss based on the Structural Similarity Index Metric (SSIM). For example, for MRA images the uPL leads to SSIM values of 0.93 while L1 and SSIM loss functions led to SSIM values of 0.81 and 0.88, respectively. 
    \EP{The} uPL network\EP{'s} initialization is not important, while network depth and pooling operations  
    impact 
    denoising performance. E.g.\ for both datasets a network with five convolutional layers led to the best performance while a network with more layers led to a performance drop. We also find that small uPL networks led to better or comparable results than using large networks such as VGG (e.g. SSIM values of 0.93 and 0.90 for a small and \EP{a} VGG19 uPL network in the MRA dataset).  We demonstrate superior performance of our loss for both datasets, all noise levels, and three network architectures. 
	In conclusion, for images containing line-like structures, uPL is an alternative to other loss functions for 3D image denoising. \EP{We observe that} 
    small uPL networks have better or equal performance than very large network architectures while requiring lower computational costs and should therefore be preferred.

	\section*{Introduction}
	Enhancing image quality of 3-dimensional (\EP{3D}) images by suppressing image noise is important in bio-medical applications. Conventional methods such as spatial filtering or other, edge-preserving techniques (e.g.\ bilateral filtering \cite{Tomasi1998,Banterle2011}) are often used for image denoising \cite{manjon2008mri, yang2015brain}. However, these conventional filtering methods tend to 
	lead to blurry results. In recent years, convolutional neural networks (CNNs) overcame the limitations of conventional denoising methods and showed strongly improved results for both \EP{2D} and \EP{3D} images \cite{zamir2022restormer, zhang2019residual}. 
	For image denoising using deep learning, several network architectures have been proposed \cite{zamir2022restormer, 9937486,chen2022nonlocal}. 
	The majority of works reporting on image denoising concentrate on the optimal network architecture. However, not only network structure but also selecting an appropriate loss function is important for network performance. 
	As the loss function is a crucial part of neural network training, the focus of this work lies in finding the optimal loss function for \EP{3D} image denoising of images containing line-like structures. \\
	
	For Magnetic Resonance (MR) images, image quality is related to scan time, i.e.\ shorter scan times result in images with higher image noise \cite{lustig2007sparse}. While causing higher noise levels, shorter scan times lead to better patient comfort and higher daily patient throughput. Image denoising can transform these shorter scans to images with similar image quality as scans acquired during a longer time. 
	
	For \EP{3D} MR image enhancement, two approaches are frequently used: (1) In the majority of the works, the images are denoised slice by slice, i.e.\ using a \EP{2D} network \cite{chung2022mr, moreno2021evaluation, koch2021analysis}. (2) Other works denoise  the whole \EP{3D} volume (\EP{3D} networks) \cite{tian2022sdndti, aetesam2022attention, ran2019denoising}. In \cite{wu2021denoising} the authors compare the \EP{2D} slice-by-slice with the pure \EP{3D} approach using various denoising networks. They demonstrate that using the \EP{3D} volume and \EP{3D} networks leads to consistently better results than the \EP{2D} approach as \EP{2D} denoising methods ignore relevant \EP{3D} information. Therefore, in this work, we concentrate on \EP{3D} networks.
	
	For \EP{2D} images, the 'Perceptual Loss' \cite{Bruna2016SuperResolutionWD, johnson2016perceptual, DosovitskiyNIPS2016, NguyenCVPR2017, DB16d, ledig2017photo, nasser2022perceptual} (also called 'content representation' \cite{Gatys_2016_CVPR}) is frequently used as loss function for image enhancement and outperformed traditional pixel-by-pixel losses. \EP{The Perceptual Loss is also often used in combination with L1- or L2-loss where it also leads to performance improvements \cite{shah2022impact}.} The standard Perceptual Loss (sPL) minimizes differences between features extracted from reconstructed and ground truth image using a pre-trained neural network. The success of the sPL has been thought to lie in the fact that the loss network extracts meaningful features, as it has been 
	trained on a large variety of images 
	\cite{ledig2017photo, johnson2016perceptual}. Such pretrained networks are unavailable in \EP{3D} hampering the application of sPL for \EP{3D} MRI enhancement. However, in recent works an untrained Perceptual Loss (uPL) was used in \EP{2D} image super-resolution and 
	compared features from untrained, randomly initialized networks \cite{liu2021generic, he2016powerful}. The randomly initialized loss networks were exclusively used to extract feature maps from high-quality images and remain constant during the super-resolution training process and performed similarly to sPL  \cite{he2016powerful}. This opens a door towards using Perceptual Losses also in \EP{3D}. 
	\\  
	For \EP{3D} image enhancement, the most frequently used loss functions are L1 or L2 loss which compare generated and ground truth images voxel-by-voxel and thereby ignoring voxel neighborhoods \cite{wu2021denoising,manjon2019mri}. In this work, we investigate if the uPL can be used beneficially for \EP{3D} images containing line-like structures. Hereby, we are interested in the performance impact of uPL network parameters. 
	We are especially interested if also small, simple networks can be used in the uPL. 
	\begin{figure} 
		\centering
		\includegraphics[width=0.59\linewidth]{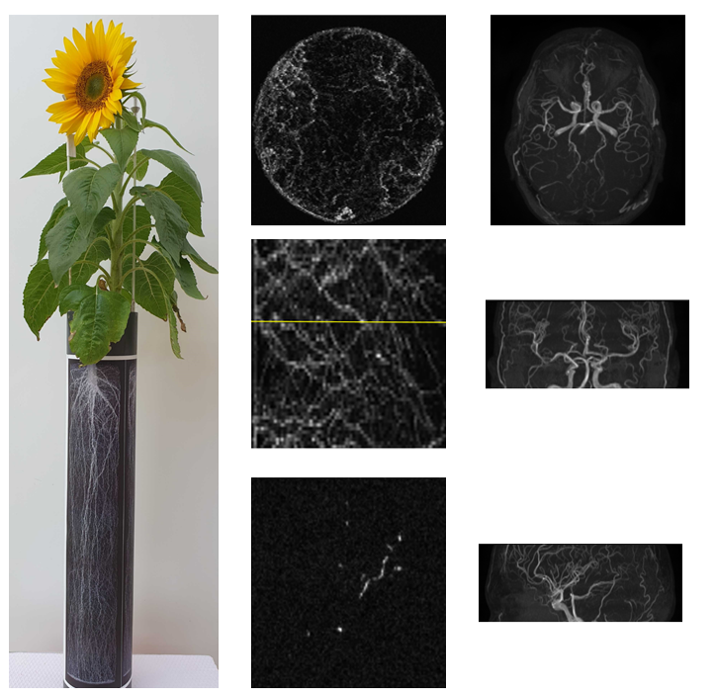}
		\caption{Example \EP{3D} MR images of plant roots and brain vessels. First column: Plant scanned in the MR system, second column: maximum intensity projections (MIP) of MR root image (top: axial plane, mid.: sagittal plane) and a single slice of the \EP{3D} image marked yellow in the sagittal plane (bot.). Next columns, top to bot.: Axial, coronal, and sagittal MIP of an MRA image.}
		\label{fig:1}
	\end{figure}
	\textbf{Our contribution:}  In summary, our contributions are as follows:
	
	\begin{itemize}%
		
		\item We propose \EP{3D} uPL with small \EP{3D} convolutional networks in the loss function for \EP{3D} denoising - a straightforward translation from the \EP{2D} case. 
		\item We demonstrate the suitability of uPL for \EP{3D} denoising of MR images containing line-like structures outperforming competing losses by a large margin.
		\item We investigate and discuss the influence of network hyper-parameters like network depth, layer thickness, kernel size, and pooling operations.
		\item  We investigate the suitability of the uPL for three denoising network architectures and demonstrate its superiority for all \EP{the three} architectures, where we also provide a novel \EP{3D} implementation of the Restormer \cite{zamir2022restormer} network.
	\end{itemize}%
	
	\section*{Materials and methods}
	\subsection*{Datasets}
	We included two datasets in this study. 
	The first dataset consists of 536 MR images of plant roots in soil with voxel size 0.5 mm $\times$ 0.5 mm $\times$ 0.99 mm and a typical image size of $192 \times 192 \times 1000$. For the root dataset, 380 images were used for training, 80 images for validation, and 76 for testing. 	
	
	The MRA data consists of MR images of brain vessels. This dataset is part of the publicly available IXI datasets \footnote{\url{https://brain-development.org/ixi-dataset/}, assessed on 09/10/2022}. We had no access to any information that could identify the scanned individuals. The typical MRA image size of this dataset is $512 \times 512 \times z_\text{dim}$. The $z_\text{dim}$ varies from patient to patient. The training set was randomly split into 300 training, 50 validation, and 100 test images. 
	
	An example image for each dataset is displayed in Fig \ref{fig:1}. For both datasets, four levels of Rician noise were artificially added to assess the impact of the loss functions for different signal-to-noise ratios (1\%, 5\%, 10\%, and 20\% noise added). Experiments regarding loss network structure were performed for noise level 3, i.e.\ 10\%. 
	
	\subsection*{Denoising networks}
	
	Three network architectures are compared in this study: a DnCNN \cite{zhang2017beyond}, a ResNet architecure \cite{ledig2017photo} and a \EP{T}ransformer network as proposed for \EP{2D} by Zamir et al.\  \cite{zamir2022restormer} translated to \EP{3D}. Please note that we also tested U-Net like architectures. However, especially for our very fine root data, these architectures generally led to much worse 
	results than the shown methods, for pixel-wise losses as well as the better but still bad performing uPL. Therefore, we concentrate in this work on the three mentioned architectures.\\ 
	The DnCNN consists of one convolutional layer with 64 feature maps and kernel size 3 followed by a ReLU activation function. Next, are  three blocks consisting of a convolutional layer with 64 feature maps and kernel size 3, a ReLU activation, and a batch normalization layer. The last layer is a convolutional layer with kernel size 3 and 1 output channel.\\
	The ResNet consists of one convolutional layer followed by a PreLU activation function. The first convolutional block is followed by five residual blocks, where each residual block contains one convolutional layer, a batch normalization layer, and a PreLU activation function. The residual blocks are connected via a skip connection. The residual blocks are followed by a convolutional layer, a batch normalization, and another PreLu activation. \\
	The \EP{T}ransformer network is specially designed such that it can be used for large images while modeling global connectivity. In this network, multi-head 'transposed' attention (MDTA) blocks are introduced applying attention across feature dimension\EP{s} rather than across spatial dimensions. \EP{Before feeding the data to the MDTA blocks, the images are resized to the size (image height $\cdot$ image width) $\times$ number of channels.} After the MDTA blocks follow feed-forward blocks consisting of two convolutional layers and a gating layer. We adjust the transposition in the attention blocks to the \EP{3D} case. \EP{In the 3D Transformer network, all 2D convolutions are replaced by 3D convolutions. Before the data is fed to the MDTA blocks, the data is resized to the size (image height $\cdot$ image width $\cdot$ image depth) $\times$ number of channels. }  As for our fine structures, UNet-like \EP{architectures} lead to performance drops, the Transformer blocks were in our study combined sequentially\EP{, in contrast to the original 2D Restormer \cite{zamir2022restormer}}. 
	A graphical overview of all network structures is displayed in the Supplemental Material (\ref{s1Fig}, \ref{s2Fig}, \ref{s3Fig}). 
 \\
	\EP{The performance of the proposed uPL is compared with the L1 loss across the mentioned network architectures and different noise levels. For the evaluation of characteristics of the uPL network, the DnCNN is used as example network. }
	\EP{\subsection*{Loss functions}
		In the next paragraphs, we first define the uPL. We then compare the denoising results using an uPL with a small loss network with conventional loss functions used in the literature. This experiment serves as a simple demonstration of the suitability of the uPL for image denoising tasks. In the next section, we then investigate the impact of different characteristics of the uPL network on the denoising results. All experiments were performed with five different random seeds. 
	}
	\EP{\subsubsection*{Untrained Perceptual Loss} 
		Perceptual Loss (PL) minimizes the differences between feature maps of neural networks of generated and ground truth images. PL is defined as 
		\begin{equation} \label{eqn}
			PL=\sum_j\frac{1}{C_{j}H_{j}W_{j}D_{j}}||\phi_{j}(\hat{y})-\phi_{j}(y)||_{2}^{2}
		\end{equation}
		where $C_{j}, H_{j}, W_{j}, D_{j}$ refer to the number of channels, the image height, the image width\EP{, and the image depth} in layer $j$, respectively. $\hat{y}$ refers to the generated image and $y$ to the corresponding ground truth image, and $\phi_{j}$ outputs the activation of the $j^\text{th}$ layer in the loss network.
	}
	
	\EP{\subsubsection*{Proof of concept - Comparison with conventional loss functions}}
	\EP{As proof of concept, we compared the results when using a small uPL network in the loss function with the most frequently used loss functions for image enhancement.}
	We compare (1) the L1 loss \cite{wu2021denoising,manjon2019mri, feng2022multi}, (2) a loss maximizing SSIM  \cite{ahn2022deep,jia2024highly}, (3) uPL with the untrained \EP{3D} version of VGG19, (4) uPL with the untrained \EP{3D} AlexNet, and (5) uPL with a simple loss network with three convolutional layers (each with 32 features and kernel size 3, ReLU activation).\\
	\EP{We use this proof of concept study and the small network as a starting point to investigate which impact different characteristics of an uPL network have on the results. }
	In particular, we concentrate on the following questions:  \textit{Can small and simple networks be successfully used in uPL? What are the important architectural parameters for successful use in uPL? What impact has the weight initialization? Is there a network that works best for all datasets or does the best network structure depend on image content? Is the impact of the uPL dependent on the denoising architecture? } 
	As there are more hyper-parameters than we can reasonably consider
	in the uPL, we concentrate on a few prominent ones. 
	Our goal is to find an easy working solution 
	rather than infeasibly covering the entire parameter space.
	We tested the performance of different uPL network characteristics for 10 \% added noise and the DnCNN architecture. If a network characteristic leads to a clear improvement, this network characteristic was used as the default characteristic. If performance across similar characteristics was similar, we used the simplest solution (i.e. the smallest network) in the consecutive tests.\vspace*{\baselineskip}\\

	\subsection*{Characteristics of networks used in uPL}
	\subsubsection*{Initialization of network used in uPL}
	As the networks in the uPL are not trained, different initializations may 
	have considerable 
	impact on the performance. 
	We investigate the impact of weight initialization in the uPL network 
	by analyzing the results using five different random seeds 
	for weight initialization in PyTorch. Weights were drawn from a uniform distribution $ \mathcal{U}(-\sqrt{k},\sqrt{k})$ where $k=(C_{in}*\prod_i K_i)^{-1}$ and $K_i$ is the size of the $i^\text{th}$ kernel. The loss network was a simple loss network with three convolutional layers, kernel size 3, and 32 feature maps. \\
	Regarding the initialization method, five initialization methods were tested: (1) Kaiming uniform, (2) Kaiming normal, (3) Xavier uniform, (4) Xavier normal, and (5) the default uniform initialization of PyTorch (i.e.\ weights are drawn from a uniform distribution) \cite{KaimingArxiv2015, XavierIni2010}. 
	
	\subsubsection*{Depth of network used in uPL}
	Previous works demonstrated that for \EP{2D} images deeper loss architectures resulted in better performance for image segmentation and super-resolution tasks \cite{ledig2017photo, liu2021generic}. We investigate if this holds also for \EP{3D} data, by varying uPL network depth. 
	We tested networks with depths 3, 5, 7, 9, and 13 convolutional layers connected with a ReLU activation function. Additionally, we tested kernel sizes of 3, 5, 7, and 9.  As default, each convolutional layer created 32 feature maps. 
	
	\subsubsection*{Impact of pooling operations of network used in uPL}
	Down-sampling a signal creates aliasing and potential loss of information when fine details or other high spatial frequency signals are present. This is why down-sampling or pooling is recommended to be applied after sufficient smoothing only (see e.g.\ \cite{ZhangCorr2019}). As noise-initialized kernels typically have poor smoothing behavior the use of pooling layers in the uPL might lead to a loss in information and therefore to a performance drop. To assess the impact of pooling layers, we use an uPL network with five convolutional layers and added max-pooling operations. We were interested if the number of pooling operations \EP{impacts} the results. The number of pooling operations varied from 1 to 3. We started with only one pooling operation after the first convolutional layer. Next, we added a second pooling layer after the second convolutional layer and a third pooling operation after the third convolutional layer. \\
	
	\subsection*{Implementation and training details}
	All experiments were implemented in Pytorch 1.10.0 and training was performed using PyTorch lightning. Networks were trained for 30.000 iterations with batch size 16, Adam optimizer \cite{Kingma2015AdamAM} and learning rate 0.001. \EP{The training parameters were chosen as they lead to the overall best performance in the validation sets.} For training, images were cropped randomly to a size of $96 \times 96 \times 96$. 
	
	\subsection*{Evaluation metrics}
	For evaluation, we calculate the structural similarity index (SSIM), peak signal-to-noise ratio (PSNR), and mean squared error (MSE) between generated and ground truth images.
	Mean \EP{and standard deviation (std) values over the datasets and random seeds are reported if not stated differently. As both datasets contain fine structures, the evaluation metrics were additionally calculated on the parts of the image containing important information. A cube of size 52 $\times$ 52 $\times$ 52 was cropped from the root data. As the roots grow from \EP{top to bottom, }
    the cube is cropped from the upper middle part of the image. A cube of size 68 $\times$ 68 $\times$ 68 is cropped from the middle of the MRA images as these images contain important information mainly in the image center. 
    An example illustration is displayed in Supplemental Fig \ref{s4Fig}. 
    } \\

	\section*{Results}\label{sec4}
	\EP{\subsection*{Proof of concept - Comparison with conventional loss functions}}
	\EP{In this experiment, the performance of different loss functions frequently used for image denoising are compared with an uPL containing a small network (three convolutional layers, kernel size 3). In this experiment, we observe consistent results for both datasets see Table~\ref{table:1}. For both datasets, uPL using a small network leads to the best evaluation metrics compared with the other loss functions included in this study. For example, uPL results in an SSIM of 0.86 $\pm$ 0.01 and a PSNR of 38.1 $\pm$ 0.4 for the roots dataset. For comparison, L1 (SSIM) loss yields a mean SSIM of 0.79 $\pm$ 0.02 (0.63 $\pm$ 0.03) and a PSNR of 37.4 $\pm$ 0.5 (31.7 $\pm$ 0.8). For the MRA dataset improvements are similar. uPL yields an SSIM of 0.9 $\pm$ 0.01 while the L1 loss yields an SSIM of 0.79 $\pm$ 0.02 and SSIM as loss function leads to an SSIM of 0.86 $\pm$ 0.04. \\
		The uPL containing a small network also outperforms the straight-forward \EP{3D} extension of uPL using AlexNet or VGG. For the MRA data, the performance drops to an SSIM of 0.86 $\pm$ 0.02 (PSNR 30.9 $\pm$ 0.3). The same holds for the root images, where the drop in performance is high. In contrast, using VGG19 in the loss leads to a slight performance drop when compared with the small network for both datasets. However, differences are not pronounced and almost not visible when comparing images by eye.} \\
	\begin{figure}[t] 
		\centering
		\includegraphics[width=0.98\linewidth]{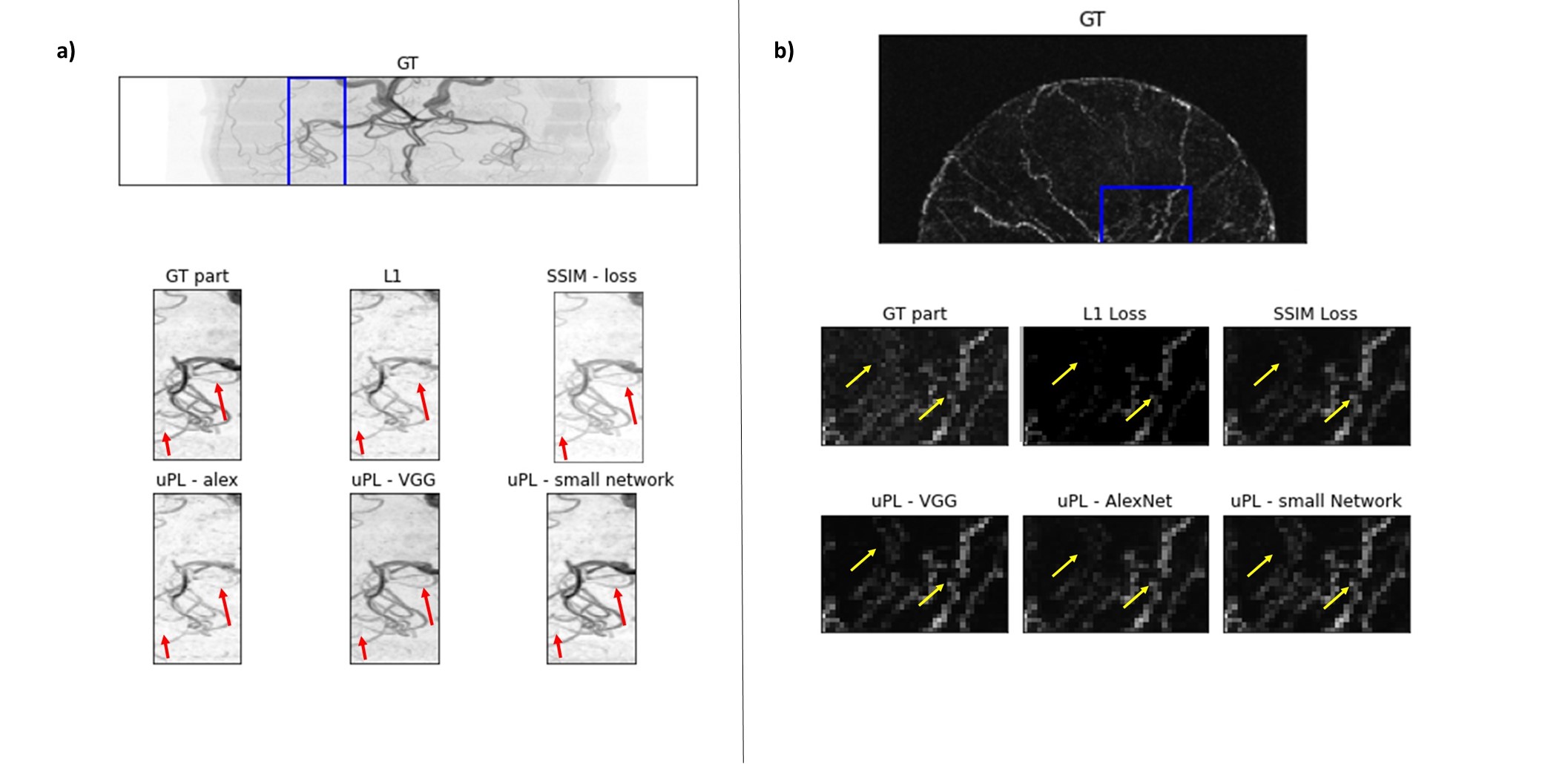}
		\caption{Example \EP{3D} MR images of plant roots and brain vessels. First column: Plant scanned in the MR system, second column: maximum intensity projections (MIP) of MR root image (top: axial plane, mid.: sagittal plane) and a single slice of the \EP{3D} image marked yellow in the sagittal plane (bot.). Next columns, top to bot.: Axial, coronal, and sagittal MIP of an MRA image.}
		\label{fig:roots}
	\end{figure}
	
	Fig ~\ref{fig:roots} shows denoising results of one MR root and one MRA image for 10\% noise added reconstructed with the DnCNN. \EP{As displayed, using L1 loss or SSIM loss leads to strong suppression of fine roots and to incorrect reconstruction of finer vessels. Also uPL using AlexNet fails to reconstruct finer roots and vessels accurately. The DnCNN trained with the smaller network (3 convolutional layers and kernel size 3), the VGG19 network creates root and MRA images where also fine details are correctly reconstructed.} 
	
	{
		\begin{table} [h!]
			\flushleft
			\resizebox{0.99\textwidth}{!}
			{
				\begin{tabular}{ || c |c c c | c c c  || }
					\hline
					&\multicolumn{3}{c|}{MRA}&\multicolumn{3}{|c||}{MR root}\\
					Loss & SSIM & PSNR & MSE  &  SSIM & PSNR & MSE\\ 
					&&&&&  & (roots)\\ [0.5 ex]
					\hline \hline
					
					L1 & \EP{0.79 $\pm$ 0.02 } & \EP{31.4 $\pm$ 0.41} & \EP{0.0047 $\pm$ 0.004 }& \EP{0.79 $\pm$ 0.02} & \EP{34.4 $\pm$ 0.5} & \EP{0.038 $\pm$ 0.006}\\
					SSIM loss & \EP{0.86 $\pm$ 0.04} & \EP{35.6 $\pm$ 1.1} & \EP{0.0051 $\pm$ 0.011} &  \EP{0.63 $\pm$ 0.03} & \EP{31.7 $\pm$ 0.8} & \EP{0.058 $\pm$ 0.008} \\
					VGG19 & \EP{0.87 $\pm$ 0.01} & \EP{40.2 $\pm$ 0.3} & \EP{0.0075 $\pm$ 0.0008} & \EP{0.84 $\pm$ 0.01}  & \EP{37.8 $\pm$ 0.3 }& \EP{0.031 $\pm$ 0.007}\\
					AlexNet & \EP{0.86 $\pm$ 0.02} & \EP{30.9 $\pm$ 0.3} & \EP{0.0043 $\pm$ 0.001} & \EP{0.75 $\pm$ 0.01} & \EP{28.4 $\pm$ 0.4} & \EP{0.043 $\pm$ 0.006}\\
					 Our SimpleNet & \EP{\underline{0.90 $\pm$ 0.01}} & \EP{\underline{41.3 $\pm$ 0.4}} & \EP{\underline{0.0043 $\pm$ 0.0004}}& \EP{\underline{0.86 $\pm$ 0.01}}  & \EP{\underline{38.3 $\pm$ 0.1}} & \EP{\underline{0.031 $\pm$ 0.005}}\\
					\hline
			\end{tabular}}
			\caption{ \EP{Mean and standard deviation values for the five random seeds used for network training for MRA images and MR root images for the loss functions included in this study. Mean and standard deviation values for the evaluation metrics calculated exclusively on the center image parts as well as std values across test sets are given in the Supplemental Information Table S2 and S3.}}
			\label {table:1}
		\end{table}

		\subsection*{Impact of network initialization}
		\EP{To demonstrate that the superior performance of our small loss network is not due to chance, we investigated the performance for five different random seeds in the loss network. Additionally, the impact of network initialization on performance metrics is investigated.} For both datasets, the differences across weight initialization seeds are small (see Supplemental Table S1 ). E.g.\ for the MRA dataset different seeds resulted in PSNR between 41.2 and 41.4 and SSIM between 0.90 and 0.91. For the root dataset, different seeds resulted in SSIM values between 0.85 and 0.86. \\
		
		\begin{table} [h!]
			\flushleft
			\resizebox{0.99\textwidth}{!}
			{
				\begin{tabular}{ || c |c c c | c c c  || }
					\hline
					uPL Loss&\multicolumn{3}{c|}{MRA}&\multicolumn{3}{|c||}{MR root}\\
					initialization & SSIM & PSNR & MSE & SSIM &  PSNR & MSE\\ 
					method&&&&&  & (roots)\\ [0.5 ex]
					\hline \hline
					Kaiming uniform & \EP{0.91 $\pm$ 0.01} & \EP{40.6 $\pm$ 0.3} & \EP{4.3e-3 $\pm$ 2e-4}& \EP{0.86 $\pm$ 0.01} & \EP{38.8 $\pm$ 0.2} & \EP{0.034 $\pm$ 0.001}\\
					Kaiming normal & \EP{0.91 $\pm$ 0.01}& \EP{40.5 $\pm$ 0.2}& \EP{4.2e-3 $\pm$ 1e-4}&  \EP{0.85 $\pm$ 0.02}& \EP{38.8 $\pm$ 0.3} &  \EP{0.034 $\pm$ 0.002}\\
					Xavier uniform & \EP{0.92 $\pm$ 0.01}& \EP{40.2 $\pm$ 0.2 }& \EP{4.0e-3 $\pm$ 2e-4} & \EP{0.85$\pm$ 0.01} &  \EP{38.8 $\pm$ 0.2} & \EP{0.032 $\pm$ 0.001}\\
					Xavier normal & \EP{0.92 $\pm$ 0.00} & \EP{40.4 $\pm$ 0.3} & \EP{4.3e-3 $\pm$ 4e-4}& \EP{0.86 $\pm$ 0.01} & \EP{38.9 $\pm$ 0.2} & \EP{0.041 $\pm$ 0.001}\\
					Default uniform & \EP{0.86 $\pm$ 0.01} & \EP{36.7 $\pm$ 0.3} &  \EP{4.9e-3$\pm$ 5e-4} &\EP{ 0.85 $\pm$ 0.01}&  \EP{38.8 $\pm$ 0.3} & \EP{0.05 $\pm$ 0.002} \\
					\hline
			\end{tabular}}
			\caption{ Mean and std evaluation metrics for MRA and MR root images for different initializations of the untrained loss network.}
			\label {table:ini}
		\end{table}
		All but the uniform initialization method le\EP{ad} to comparable results across datasets (see Table ~\ref{table:ini}). For the MRA dataset, uniform initialization lead to \EP{SSIM performance drop} from e.g.\ 0.92 for Xavier normal to 0.86 to uniform initialization. For the root dataset, no difference between initialization methods can be observed. We therefore use the \EP{Xavier normal} initialization in all other experiments.  
		
		\subsection*{Number of convolutional layers vs.\ kernel size}
		
		\begin{table} [h]
			\centering
			{
				\begin{tabular}{ || c | c c c  c  || }
					\hline
					Metric&\multicolumn{4}{ c ||}{Number of conv. layers}\\ \hline
					
					Kernel size & 3 conv & 5 conv &   9 conv & 13 conv \\ [0.5 ex]
					\hline
					\hline
					&\multicolumn{4}{c||}{SSIM - MRA dataset}\\
					\hline \hline
					3 &  \EP{0.90 $\pm$ 0.01} & \EP{0.93$\pm$ 0.01} & \EP{0.90$\pm$ 0.02} & \EP{0.90$\pm$ 0.01}  \\
					5 &\EP{0.92 $\pm$ 0.01} & \EP{0.93$\pm$ 0.01} & \EP{0.90$\pm$ 0.01} & \EP{0.90$\pm$ 0.01}  \\
					7 & \EP{0.93 $\pm$ 0.00} & \EP{0.86 $\pm$ 0.01} & \EP{0.90$\pm$ 0.01} & \EP{0.90$\pm$ 0.01}\\
					9 & \EP{0.92 $\pm$ 0.01} & \EP{0.91 $\pm$ 0.01} & \EP{0.89$\pm$ 0.02} & \EP{0.89$\pm$ 0.02}\\
					
					\hline \hline
					&\multicolumn{4}{c||}{SSIM - Root dataset}\\
					\hline 
					3 & \EP{0.86 $\pm$ 0.01 } & \EP{0.83 $\pm$ 0.01} & \EP{0.84 $\pm$ 0.02} & \EP{0.84 $\pm$ 0.03} \\
					5 & \EP{0.85 $\pm$ 0.01 } & \EP{0.86 $\pm$ 0.02 } & \EP{0.84 $\pm$ 0.02 } & \EP{0.83 $\pm$ 0.02 } \\
					7 & \EP{0.84 $\pm$ 0.01 } & \EP{0.84 $\pm$ 0.01 } & \EP{0.84 $\pm$ 0.02 } & \EP{0.83  $\pm$ 0.03 } \\
					9 & \EP{0.84 $\pm$ 0.01 } & \EP{0.83 $\pm$ 0.01 } & \EP{0.85  $\pm$ 0.01} & \EP{0.84  $\pm$ 0.02}\\
					\hline
			\end{tabular}}
			\caption{Mean and std SSIM values for different kernel sizes and network depth for MRA (above) \EP{and} for the MR root dataset (below). All other evaluation metrics can be found in the Supplemental Tables S4 and S5.}
			\label{table:2}
		\end{table}
		
		Results in Table \ref{table:2} show \EP{that} the number of convolutional layers and kernel sizes had a minor impact on the results. \\
		For the MRA images, \EP{seven convolutional layers with kernel size 3}, five convolutional layers with kernel size 3 or 5 lead with an SSIM of 0.93 to the best results.  Similarly, for the root dataset, five convolutional layers with kernel size 5 led with an SSIM of 0.86 to the best results, followed by three convolutional layers with kernel size 3. \EP{Taking the variance over the five random seeds into account, the difference across numbers of layers and kernel size is small.} The kernel size has little impact on the results for both dataset. A smaller network depth is beneficial. 
		
		\subsection*{Impact of pooling layers}
		\begin{figure}[t]

			\includegraphics[width=0.98\linewidth]{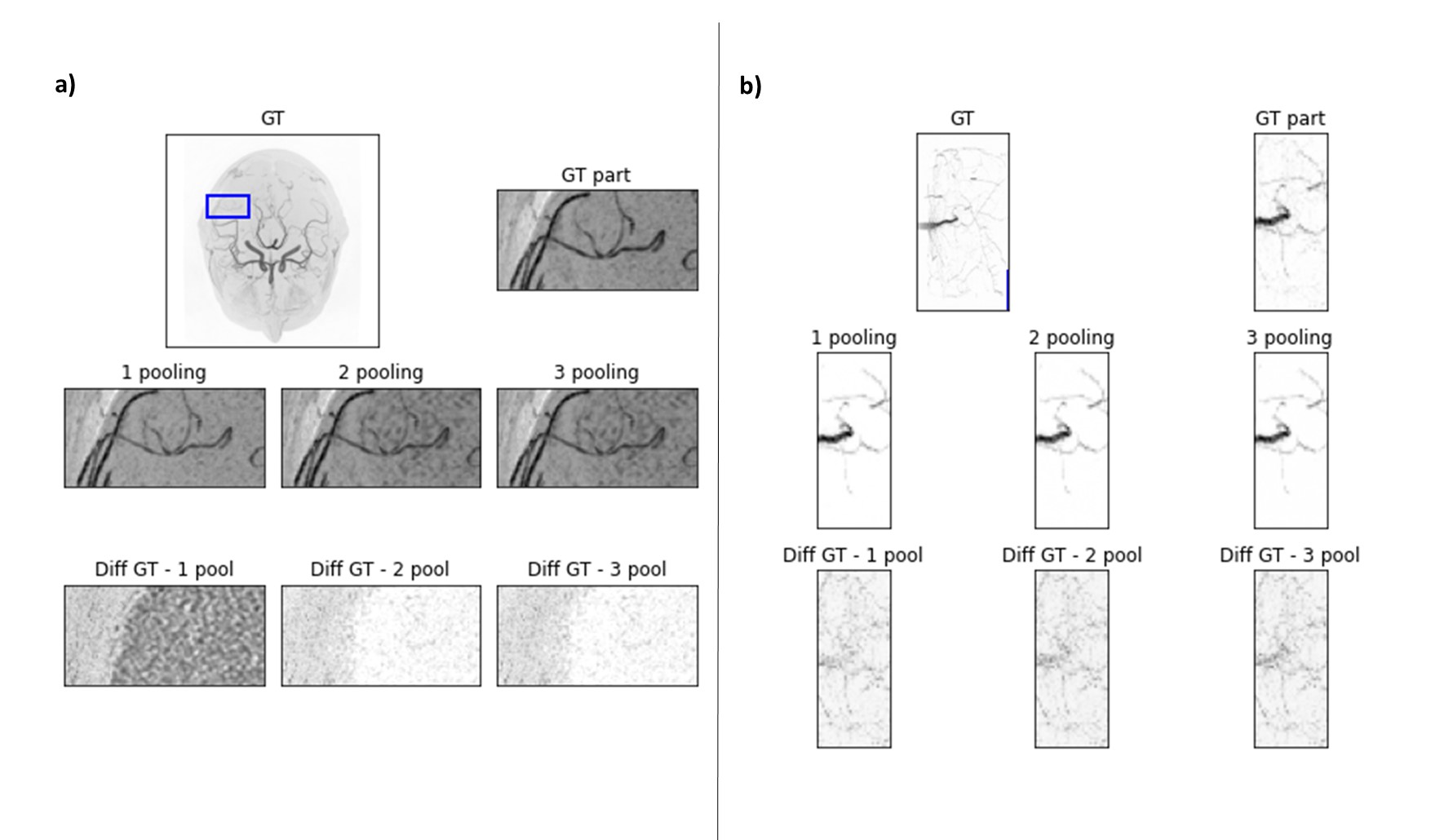}\\

			\caption{Impact of pooling operations in the uPL on the results (Left: MRA image, Right: MR Root image). First row: GT image and part of GT image. Second row: Reconstructed image with different number of pooling operations in the uPL network. Third row: Difference image between GT and reconstructed image.}
			\label{fig:pool}
		\end{figure}
		\EP{As pooling operations reduce image details, we investigate if the use of pooling operations might have an impact on the results.} For the root images, pooling operations hurt performance sligh\EP{tl}y (e.g. SSIM without pooling 0.87, with one max pooling operation 0.85). 
		For the MRA dataset, max pooling operations \EP{increase} evaluation metrics for 1 and 2 pooling operations (SSIM 0.96 and 0.94). However, evaluation metrics dropped slightly for 3 pooling operations (SSIM 0.91). This impact is illustrated in Fig \ref{fig:pool}.

		\subsection*{Denoising network architecture vs.\ loss function for different noise levels}
		
		Lastly, we investigated the impact of the denoising network architecture \EP{across} noise levels. Hereby, we were interested \EP{in the benefits of uPL across network architectures}. We investigated the performance for the previously described DnCNN, ResNet, and Transformer architectures across noise levels. We used a network with three convolutional layers and kernel size 3 in the uPL. 
		
		\begin{table} [tbh!]
			\centering
			{
				\begin{tabular}{ || c | c c c  c  || }
					\hline
					&\multicolumn{4}{ c ||}{SSIM - MR root dataset}\\ \hline
					
					Network/Loss & 1 \% noise  &   5 \% noise & 10 \% noise & 20 \% noise\\ 
					\hline \hline
					DnCNN/L1&  \EP{0.83 $\pm$ 0.02  } & \EP{0.81 $\pm$ 0.01} & \EP{0.79 $\pm$ 0.02 } & \EP{0.78 $\pm$ 0.01}  \\
					DnCNN/uPL&  \EP{\underline{0.86 $\pm$ 0.01}} & \EP{\underline{0.86 $\pm$ 0.01}} & \EP{\underline{0.86 $\pm$ 0.01}} & \EP{\underline{0.82 $\pm$ 0.0}}  \\
					ResNet/L1 &  \EP{0.83 $\pm$ 0.02} & \EP{0.83 $\pm$ 0.02} & \EP{0.83 $\pm$ 0.02} & \EP{0.80 $\pm$ 0.01}  \\
					ResNet/uPL &  \EP{\underline{0.86 $\pm$ 0.01}} & \EP{\underline{0.85 $\pm$ 0.01}} & \EP{\underline{0.84 $\pm$ 0.01}} & \EP{\underline{0.82 $\pm$ 0.0}}  \\
					
					Transformer/L1 &  \EP{0.77 $\pm$ 0.02 } & \EP{0.79 $\pm$ 0.01} & \EP{0.78 $\pm$ 0.02} & \EP{0.72 $\pm$ 0.03}  \\
					Transformer/uPL &  \EP{0.81 $\pm$ 0.01} & \EP{\underline{0.82 $\pm$ 0.01}} & \EP{\underline{0.82 $\pm$ 0.0}} & \EP{\underline{0.81 $\pm$ 0.01}}  \\

					\hline
					\hline 
					&\multicolumn{4}{c||}{SSIM - MRA dataset}\\
					\hline 
					DnCNN/L1&  \EP{0.86 $\pm$ 0.01 } & \EP{0.81 $\pm$ 0.01 } & \EP{0.79 $\pm$ 0.02} & \EP{0.81 $\pm$ 0.01}  \\
					DnCNN/uPL&  \EP{\underline{0.92 $\pm$ 0.01}} & \EP{\underline{0.90 $\pm$ 0.01 }} & \EP{\underline{0.87 $\pm$ 0.01 }} & \EP{\underline{0.84 $\pm$ 0.02 }}  \\
					ResNet/L1 &  \EP{\underline{0.99 $\pm$ 0.01 }} & \EP{0.87 $\pm$ 0.02 } & \EP{0.81 $\pm$ 0.01 } & \EP{\underline{0.84 $\pm$ 0.01 }}  \\
					ResNet/uPL &  \EP{0.98 $\pm$ 0.01 } & \EP{\underline{0.92 $\pm$ 0.01 }} & \EP{\underline{0.85 $\pm$ 0.01}} & \EP{0.82 $\pm$ 0.02 } \\
					
					Transformer/L1 &  \EP{0.97 $\pm$ 0.03 } &\EP{0.91 $\pm$ 0.02 } & \EP{\underline{0.85 $\pm$ 0.02 }} & \EP{0.68 $\pm$ 0.03 }  \\
					Transformer/uPL & \EP{\underline{ 0.98 $\pm$ 0.02}} & \EP{\underline{0.93  $\pm$ 0.01}} & \EP{0.83 $\pm$ 0.01 } & \EP{\underline{0.78 $\pm$ 0.02 }}  \\
					
					\hline
					\hline 
					
					\hline
			\end{tabular}}
			\caption{\EP{ Mean and std SSIM values across different random seeds for both datasets calculated.  All other evaluation metrics, std values across the testsets, and evaluation metrics calculated on image parts only are listed in the Supplemental Material (Table S6 - S10).}}
			\label{table:noise}
		\end{table}

		Table \ref{table:noise} \EP{lists the} SSIM for both datasets and \EP{all} noise levels. 
		The improvement when using uPL \EP{depends} on the network architecture and the noise level. Occasionally, uPL performed equal or slightly worse than L1 loss. For the \EP{T}ransformer network, the uPL showed its benefit, especially for higher noise levels (SSIM of \EP{0.78 $\pm$ 0.02 (0.81 $\pm$ 0.01) for uPL and 0.68 $\pm$ 0.03 (0.72 $\pm$ 0.03)} for L1 for the MRA (root) dataset and 20 \% noise).  The benefit of uPL was similar for DnCNN and ResNet architectures.\\
		\begin{figure}[t] 
			\centering
			\includegraphics[width=0.99\linewidth]{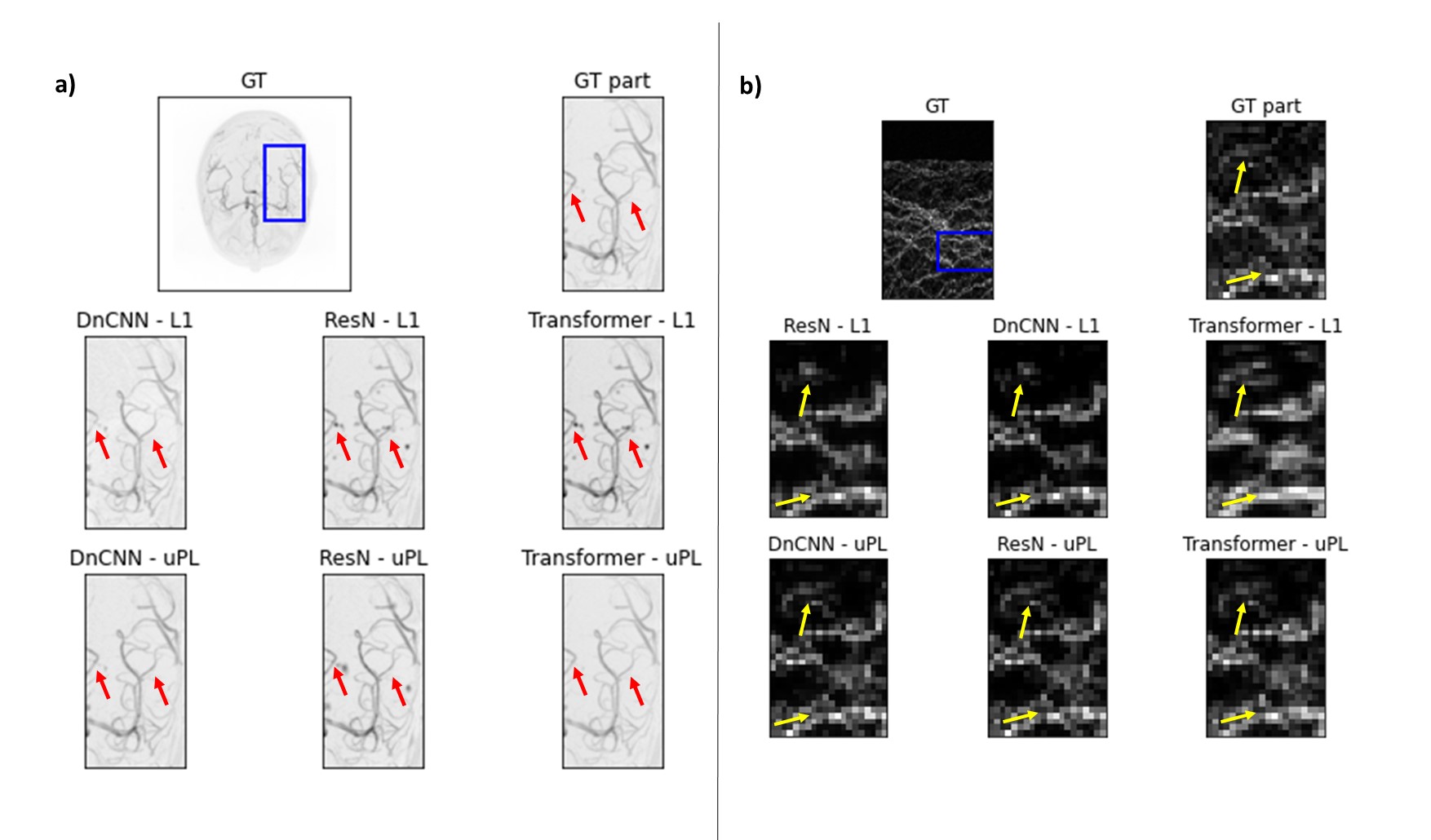}
			\caption{Example \EP{3D} MR images of plant roots and brain vessels. First column: Plant scanned in the MR system, second column: maximum intensity projections (MIP) of MR root image (top: axial plane, mid.: sagittal plane) and a single slice of the \EP{3D} image marked yellow in the sagittal plane (bot.). Next columns, top to bot.: Axial, coronal, and sagittal MIP of an MRA image.}
			\label{fig:2}
		\end{figure}
		As expected, denoising performance was best for the lowest noise levels: The uPL resulted in SSIM values of 0.82 - 0.86 for the roots data and of 0.9 - 0.99 for the MRA data. The L1 loss led to SSIM values between 0.81 - 0.84 for the root and 0.81 - 0.98 for the MRA data. The uPL outperformed especially in higher noise levels the L1 loss although the denoising performance decreases for higher noise levels. \EP{For the MRA data, the superiority in performance was not visible for the lowest and highest noise level and the ResNet, and 10 \% added noise and the Transformer network. }

		\section*{Discussion}\label{sec5}
		In summary, we demonstrated that the untrained Perceptual Loss leads to clear improvement for image denoising of MR images displaying line-like structures. \\
		We observe that for both datasets, the uPL leads especially with small uPL networks to very good results. 
		We demonstrated that loss network initialization has a minor impact on the results. However, more parameters might be important and future research may find better loss networks. For example, the use of other non-linearities than ReLU might influence the results and introducing layer weights or masks may be pertinent (similar to the 'diversity loss' in \cite{Chen_2017_ICCV}). We plan to address these points in future work.\\   
		It is hard to compare our results directly with results from other works due to the special characteristics of our datasets. Especially the root dataset is very sparse and yields very thin structures. As in our experiments, U-Net-like architectures failed for the MR root images, other famous denoising networks were not tested in this study \cite{gurrola2021residual, chen2022hider}. We hypothesize that the down-sampling part of a U-Net architecture leads to a loss of information as the roots are very fine structures. We found only a small number of works that report on image denoising of MR angiographs. In \cite{xue2024darcs} the authors used also an U-Net-shaped architecture. However, by testing three different denoising networks - one of them a Transformer network, we aimed to demonstrate that the success of the uPL is not network-dependent and can improve the results of different network architectures. \\
		
		In this work, we focused on denoising of MR images displaying line-like structures. These images contain very fine details and yield therefore special challenges. In future work, we will investigate if the uPL can also be used for medical images yielding other image characteristics such as MR images of the human brain, PET or CT images.\\ 
		Especially for difficult tasks, i.e.\ with increasing noise level the uPL performs considerably better than L1 loss for both datasets. This is probably because the uPL is taking neighborhood information into account \EP{by using convolutional kernels with sizes larger than 1} when calculating the loss. Therefore, even if one pixel is highly disturbed by noise, information about the ground truth pixel can still be estimated by its neighbors. \\
		Regarding network architecture, we observed similar or even better performance when using uPL and a DnCNN than when combining L1 loss and one of the other network structures. These results indicate that the use of uPL makes it possible to use computationally less expensive networks without a performance drop. Additionally, our results suggest that the additional value of a Transformer network is not visible for both datasets. This is in contrast to other works that showed superior performance for similar denoising \EP{T}ransformers \cite{bai2024mrformer}. The lower performance of the \EP{T}ransformer network is likely due to the sparsity of the MR root and MRA images. Locally, both datasets contain a rather low variety of different features, i.e.\ 'just' lines and not rich texture. Therefore, large, computationally expensive, and complex networks such as the tested Transfomer variant may tend to overfit. \\
		\EP{The success of perceptual loss was thought to be due to the features networks learned during training. However, as previous studies demonstrated for 2D images, pre-trained and untrained networks used in the loss function lead to comparable results \cite{liu2021generic, he2016powerful}. This might be because randomly initialized weights can also represent the statistical properties of the training data. The pre-trained networks are usually trained on classification tasks. However, other image characteristics could be more important for image denoising of an entire image. E.g. the display of sharp edges can also be captured by convolutions and untrained feature maps. Additionally, the perceptual distance between two images is equivalent to the maximum mean discrepancy (MMD) distance between local distributions of small patches in the two images \cite{amir2021understanding}. As also demonstrated in this study, different network structures in the loss function have a high impact on the denoising results. These findings suggest that the design of the network architecture used in the loss function can be chosen such that it captures the most important image information.}
		With a best mean SSIM of 0.87 the denoising results for our root dataset still need improvement to allow for reduced measurement time in the targeted application. We observed that small roots are strongly suppressed already by low amounts of noise. None of the tested denoising network/loss function combinations allow to recover these fine details, even though the performance increase using our uPL is considerable. In consequence, as of now, plants containing very fine roots need to be scanned for a longer time than plants with thick roots.\\
		
		However, our results demonstrate the benefit of the untrained Perceptual Loss for both 3D datasets and all noise levels.
		
		\section*{Acknowledgments}
		We acknowledge the help of Dagmar van Dusschoten in providing MR images. This work was supported by the President’s Initiative and Networking Funds of the Helmholtz Association of German Research Centres [Grant HighLine ZT-I-PF-4-042]. The authors gratefully acknowledge the computing time granted through	JARA on the supercomputer JURECA \cite{thornig2021jureca} at Forschungszentrum Jülich.

	\vspace*{0.2in}
	
	\begin{flushleft}
		{\Large
			\textbf\newline{Supplemental information - Untrained Perceptual Loss for image denoising of line-like structures in MR images} 
		}
		\newline

	\end{flushleft}

		\section*{Supporting information}
		\setcounter{figure}{0}
		\setcounter{table}{0}
		\renewcommand{\figurename}{}
		\renewcommand{\tablename}{}
		\subsection*{Illustration of denoising networks}
		\label{S1_Fig}
		\renewcommand{\thefigure}{S1 Fig}
		\begin{figure} [ht]
			\captionsetup[subfigure]{justification=centering}
			\centering
			
			\includegraphics[width=0.79\linewidth]{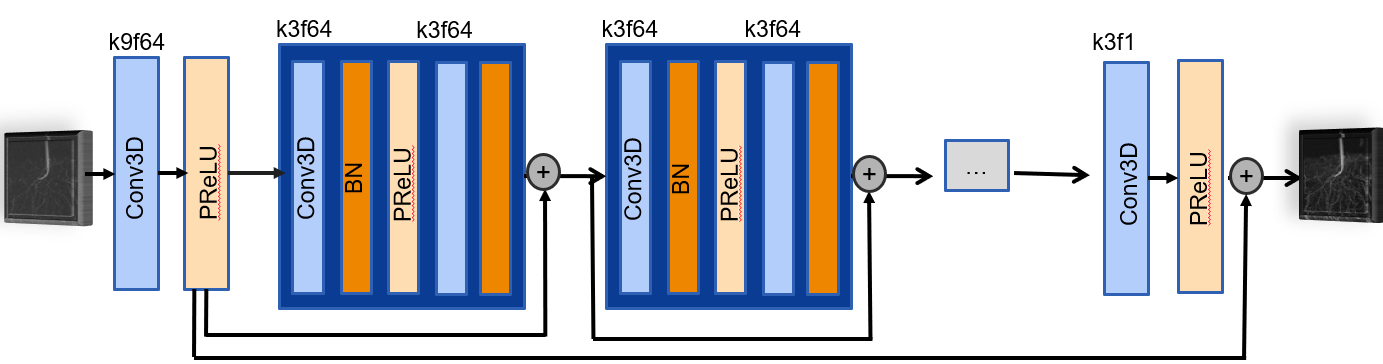}\\
			\caption{Illustration of the denoising network ResNet. The residual blocks are repeated five times. The first convolutional layer yields kernel size 9 with 64 output channels. All other convolutional layers yield kernel size 3 and 64 output channels. The last convolutional layer has kernel size 3 and 1 output channel.}
			\label{s1Fig}
		\end{figure}
		\renewcommand{\thefigure}{S2 Fig}
		\begin{figure} [ht]
			\includegraphics[width=0.79\linewidth]{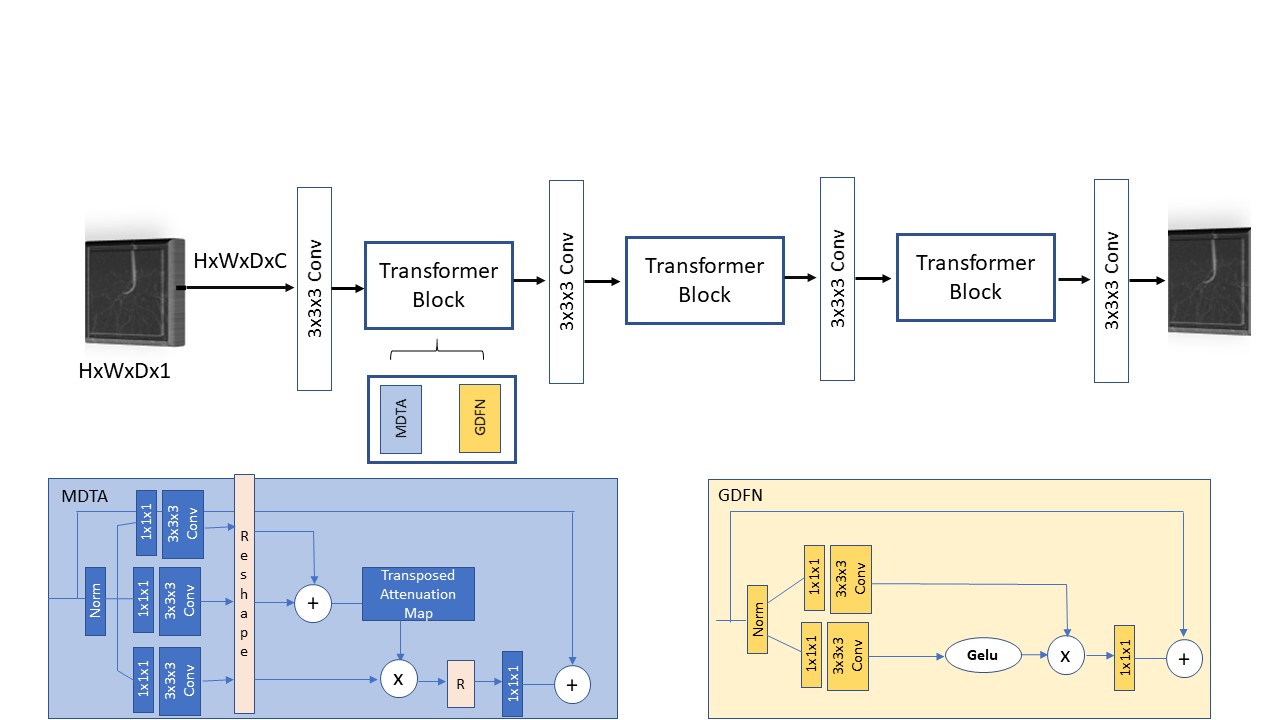}\\
			\caption{Illustration of the denoising Transformer. The Transformer blocks are organized sequentially. Details about the \EP{T}ransformer blocks are explained in the corresponding paper for 2D.}
			\label{s2Fig}
		\end{figure}
		\renewcommand{\thefigure}{S3 Fig}
		\begin{figure} [ht]
			\includegraphics[width=0.79\linewidth]{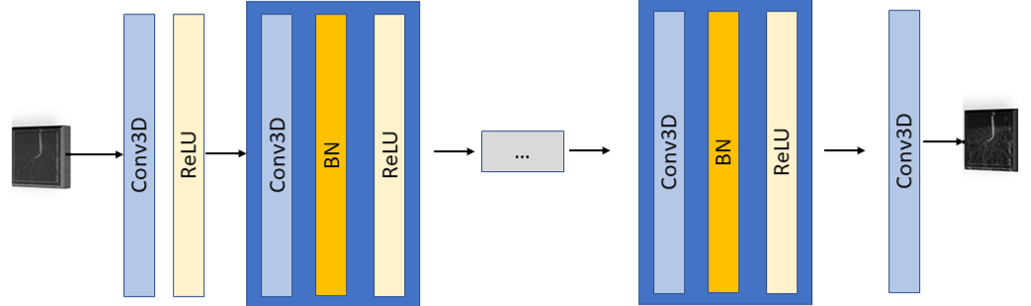}\\
			\caption{Illustration of the denoising network \EP{DnCNN}. }
			
			
		\label{s3Fig}

	\end{figure}
	
	\newpage

		\renewcommand{\thefigure}{S4 Fig}
		\begin{figure} [ht]
			\includegraphics[width=0.79\linewidth]{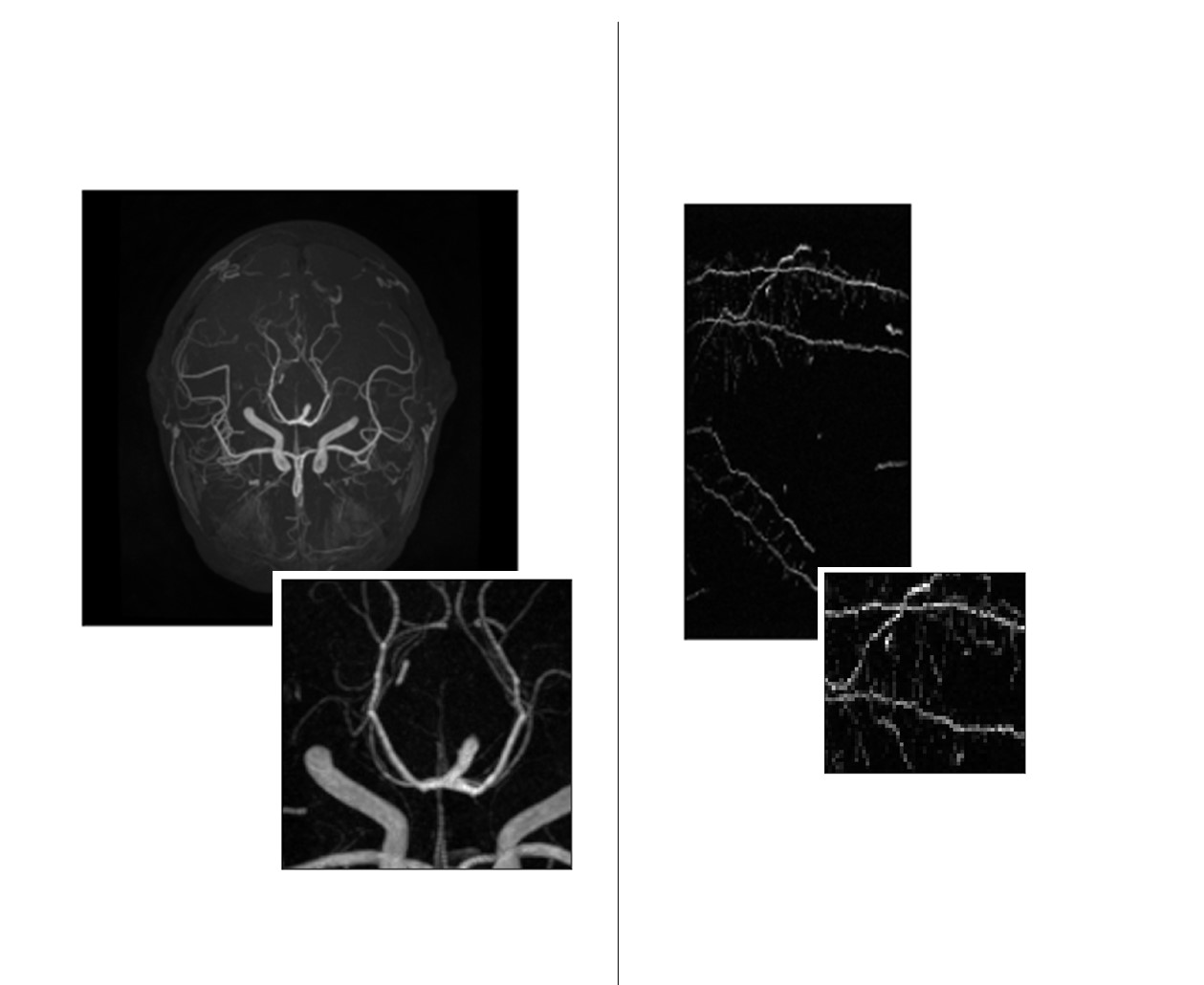}\\
			\caption{Illustration of cropped image parts on which evaluation metrics were calculated.}
			
			
		\label{s4Fig}

	\end{figure}

	\newpage
	\subsection*{Supporting Tables}
	\renewcommand{\thetable}{S1 Table}
 \begin{table} [ht]
		\centering
		{
			\begin{tabular}{ || c |c c c | c c c  || }
				\hline
				&\multicolumn{3}{c|}{MRA}&\multicolumn{3}{|c||}{MR root}\\
				Seed & SSIM & PSNR & MSE & SSIM &  PSNR & MSE\\ 
				&&&&& & (roots)\\ [0.5 ex]
				\hline \hline
				
				1& 0.89 & 32.1 & \EP{4.2e-3} & 0.83 & 37.7 & 0.033\\
				2 & 0.90 & 31.9& \EP{4.1e-3}&  0.84 & 37.8 &  0.032\\
				3 & 0.91 & 32.0 & \EP{4.0e-3} & 0.84 &  37.9 & 0.032\\
				4 & 0.89 & 32.5 & \EP{4.3e-3} & 0.83  & 37.8 & 0.032\\
				5 & 0.90& 32.5&  \EP{4.3e-3} & 0.84 &  37.8 & 0.032 \\
				\hline
		\end{tabular}}
		\label{s_table1}
		\caption{Evaluation metrics for different random seeds for MRA and root dataset.}
	\end{table}
		\renewcommand{\thetable}{S2 Table}
 \begin{table} [ht]
			\flushleft
			\resizebox{0.98\textwidth}{!}
			{
				\begin{tabular}{ || c |c c c | c c c  || }
					\hline
					&\multicolumn{3}{c|}{MRA}&\multicolumn{3}{|c||}{MR root}\\
					Loss & SSIM & PSNR & MSE  &  SSIM & PSNR & MSE\\ 
					&&&&&  & (roots)\\ [0.5 ex]
					\hline \hline
					
					L1 & \EP{0.83 $\pm$ 0.01} & \EP{30.2 $\pm$ 0.4} & \EP{0.0031 $\pm$ 6e-4}& \EP{0.73 $\pm$ 0.01} & \EP{34.2 $\pm$ 0.3} & \EP{0.012 $\pm$ 1e-3}\\
					SSIM loss & \EP{0.82 $\pm$ 0.02} & \EP{31.1 $\pm$ 0.8} & \EP{0.030 $\pm$ 0.009} &  \EP{0.74 $\pm$ 0.1} & \EP{34.0 $\pm$ 0.7} & \EP{0.067 $\pm$ 0.008} \\
					VGG19 & \EP{0.86 $\pm$ 0.02} & \EP{33.8 $\pm$ 0.2} & \EP{0.0021 $\pm$ 0.0007} & \EP{0.74 $\pm$ 0.1}  & \EP{35.19 $\pm$ 0.4 }& \EP{0.025 $\pm$ 0.006}\\
					AlexNet & \EP{0.84 $\pm$ 0.02} & \EP{30.9 $\pm$ 0.3} & \EP{0.0033 $\pm$ 0.001} & \EP{0.54 $\pm$ 0.3} & \EP{25.4 $\pm$ 0.6} & \EP{0.051 $\pm$ 0.005}\\
					SimpleNet & \EP{0.87 $\pm$ 0.01} & \EP{33.9 $\pm$ 0.3} & \EP{0.017 $\pm$ 7e-4}& \EP{0.77 $\pm$ 0.1}  & \EP{35.5 $\pm$ 0.3} & \EP{0.0023 $\pm$ 7e-4}\\
     
					\hline
			\end{tabular}}
			\caption{ Evaluation metrics for MRA images and MR root images \EP{for the loss functions included in this study calculated on image parts only. Given are the mean and standard deviation values for the five random seeds used for network training.}.}
			\label{table:IP1}
		\end{table}

		\renewcommand{\thetable}{S3 Table}
	\begin{table} [ht]
			\flushleft
			\resizebox{0.99\textwidth}{!}
			{
				\begin{tabular}{ || c |c c c | c c c  || }
					\hline
					&\multicolumn{3}{c|}{MRA}&\multicolumn{3}{|c||}{MR root}\\
					Loss & SSIM & PSNR & MSE  &  SSIM & PSNR & MSE\\ 
					&&&&&  & (roots)\\ [0.5 ex]
					
					\hline \hline
					L1 & \EP{0.79 $\pm$ 0.06 } & \EP{31.4 $\pm$ 1.3} & \EP{0.0047 $\pm$ 0.002 }& \EP{0.79 $\pm$ 0.07} & \EP{37.4 $\pm$ 1.5} & \EP{0.038 $\pm$ 0.018}\\
					SSIM loss & \EP{0.86 $\pm$ 0.05} & \EP{35.6 $\pm$ 1.4} & \EP{0.0051 $\pm$ 0.011} &  \EP{0.63 $\pm$ 0.03} & \EP{31.7 $\pm$ 1.2} & \EP{0.058 $\pm$ 0.021} \\
					VGG19 & \EP{0.87 $\pm$ 0.05} & \EP{40.2 $\pm$ 1.9} & \EP{0.0075 $\pm$ 0.012} & \EP{0.84 $\pm$ 0.07}  & \EP{37.8 $\pm$ 1.8 }& \EP{0.031 $\pm$ 0.013}\\
					AlexNet & \EP{0.86 $\pm$ 0.06} & \EP{30.9 $\pm$ 1.7} & \EP{0.0043 $\pm$ 0.027} & \EP{0.75 $\pm$ 0.05} & \EP{28.4 $\pm$ 1.3} & \EP{0.043 $\pm$ 0.023}\\
					 Our SimpleNet & \EP{0.90 $\pm$ 0.04} & \EP{41.3 $\pm$ 0.9} & \EP{0.0043 $\pm$ 0.0013}& \EP{0.86 $\pm$ 0.05}  & \EP{38.3 $\pm$ 0.8} & \EP{0.021 $\pm$ 0.012}\\
					\hline
			\end{tabular}}
			\caption{ Mean and std values calculated over the testset for one random seed. Values are similar across random seeds.  }
			\label{table:STDT}
		\end{table}
	
	\renewcommand{\thetable}{S4 Table}
	\begin{table} [ht]
		\centering
		{
			\begin{tabular}{ || c |  c c  c  c || }
				\hline
				Metric&\multicolumn{4}{ c ||}{MSE (roots) - MR root dataset}\\ \hline
				&\multicolumn{4}{ c ||}{Number of
					convolutional layers}\\ 
				Kernel size & 3 conv & 5 conv &   9 conv & 13 conv \\ [0.5 ex]
				\hline \hline
				
				3 &  \EP{0.031 $\pm$ 0.005} & \EP{0.035 $\pm$ 0.004} & \EP{0.034 $\pm$ 0.004} & \EP{0.032 $\pm$ 0.004}  \\
				5 & \EP{0.035 $\pm$ 0.004} & \EP{0.038 $\pm$ 0.005} & \EP{0.035 $\pm$ 0.005} & \EP{0.033 $\pm$ 0.005}  \\
				7 & \EP{0.033 $\pm$ 0.006} & \EP{0.033 $\pm$ 0.005} & \EP{0.033 $\pm$ 0.006} & \EP{0.034 $\pm$ 0.006}\\
				9 & \EP{0.033 $\pm$ 0.005} & \EP{0.036 $\pm$ 0.006} & \EP{0.033 $\pm$ 0.006} & \EP{0.034 $\pm$ 0.006}\\
				
				\hline \hline
				&\multicolumn{4}{c||}{MSE - MR root dataset}\\
				\hline 
				1 & \EP{7.39e-6 $\pm$ 7e-7 } & \EP{7.89e-6 $\pm$ 6e-7} & \EP{9.31e-6 $\pm$ 7e-7} & \EP{9.12e-6 $\pm$ 8e-7} \\
				3 & \EP{7.39e-6 $\pm$ 6e-7} & \EP{5.36e-6 $\pm$ 7e-7} & \EP{8.91e-6 $\pm$ 6e-7} & \EP{9.31e-6 $\pm$ 8e-7} \\
				5 & \EP{6.41e-6 $\pm$ 8e-7} & \EP{5.75e-6 $\pm$ 8e-7} & \EP{5.67e-6 $\pm$ 8e-7} & \EP{6.12e-6 $\pm$ 7e-7} \\
				7 & \EP{5.68e-6 $\pm$ 7e-7} & \EP{5.93e-6 $\pm$ 7e-7} & \EP{6.71e-6 $\pm$ 7e-7}   & \EP{6.51e-6 $\pm$ 9e-7}\\
				9 & \EP{6.52e-6 $\pm$ 7e-7} & \EP{6.12e-6 $\pm$ 6e-7} & \EP{7.41e-6 $\pm$ 8e-7} & \EP{7.56e-6 $\pm$ 9e-7} \\
				\hline
		\end{tabular}}
		\caption{\EP{MSE for different kernel sizes and network depth: MSE calculated only for roots regions (above) and MSE for the whole image also for the MR root dataset (below).}}
		\label{s_table2}
	\end{table}
	\renewcommand{\thetable}{S5 Table}
	\begin{table} [ht]
		\centering
		{
			\begin{tabular}{ || c | c c  c  c || }
				\hline
				Metric&\multicolumn{4}{ c ||}{PSNR - MRA dataset}\\ \hline
				&\multicolumn{4}{ c ||}{Number of
					convolutional layers}\\ 
				Kernel size & 3 conv & 5 conv &   9 conv & 13 conv \\ [0.5 ex]
				\hline \hline
				3 &  \EP{41.3 $\pm$ 0.2} & \EP{41.2 $\pm$ 0.2} & \EP{40.0 $\pm$ 0.3} & \EP{40.7 $\pm$ 0.4}  \\
				5 & \EP{41.1 $\pm$ 0.3} & \EP{41.6 $\pm$ 0.3} & \EP{39.3 $\pm$ 0.4} & \EP{40.0 $\pm$ 0.3}  \\
				7 & \EP{40.2 $\pm$ 0.3} & \EP{39.4 $\pm$ 0.4} & \EP{38.4 $\pm$ 0.3} & \EP{39.5 $\pm$ 0.5}\\
				9 & \EP{41.1 $\pm$ 0.3} & \EP{39.2 $\pm$ 0.3} & \EP{40.3 $\pm$ 0.4} & \EP{39.5 $\pm$ 0.5}\\
				
				\hline \hline
				&\multicolumn{4}{c||}{MSE - MRA dataset}\\
				\hline 
				3 & \EP{0.0043 $\pm$ 0.0004 } & \EP{0.0047 $\pm$ 0.0004} & \EP{0.0072 $\pm$ 0.0005} & \EP{0.0071 $\pm$ 0.0006}  \\
				5 & \EP{0.0051 $\pm$ 0.0006} & \EP{0.0083 $\pm$ 0.0004} & \EP{0.0041 $\pm$ 0.0004} & \EP{0.0041 $\pm$ 0.0003}  \\
				7 & \EP{0.0043 $\pm$ 0.0005} & \EP{0.0088 $\pm$ 0.0006} & \EP{0.0052 $\pm$ 0.0005} & \EP{0.0051 $\pm$ 0.0004}  \\
				9 & \EP{0.0042 $\pm$ 0.0004} & \EP{0.0071 $\pm$ 0.0005} & \EP{0.0046 $\pm$ 0.0006} & \EP{0.0045 $\pm$ 0.0004} \\
				\hline
		\end{tabular}}
		\caption{\EP{PSNR/MSE} for different kernel sizes and network depth for MRA (above) and MSE for the MRA dataset (below). }
		\label{s_table3}
	\end{table}
	
	\renewcommand{\thetable}{S6 Table}
	\begin{table} [tbh!]
		\centering
		{
			\begin{tabular}{ || c | c c c  c  || }
				\hline
				&\multicolumn{4}{ c ||}{PSNR - MR root dataset}\\ \hline
				
				Network/Loss & 1 \% noise  &   5 \% noise & 10 \% noise & 20 \% noise\\ 
				\hline \hline
				DnCNN/L1&  \EP{36.3 $\pm$ 0.3} & \EP{35.0 $\pm$ 0.3} & \EP{34.4 $\pm$ 0.5} & \EP{34.0 $\pm$ 0.4}  \\
				DnCNN/uPL&  \EP{37.2 $\pm$ 0.2} & \EP{35.9 $\pm$ 0.3} & \EP{35.3 $\pm$ 0.2} & \EP{34.8 $\pm$ 0.3}  \\
				ResNet/L1 &  \EP{36.2 $\pm$ 0.1} & \EP{35.1 $\pm$ 0.2} & \EP{35.0 $\pm$ 0.2} & \EP{34.7 $\pm$ 0.2}  \\
				ResNet/uPL &  \EP{36.4 $\pm$ 0.3} & \EP{36.0 $\pm$ 0.3} & \EP{35.7 $\pm$ 0.3} & \EP{35.0 $\pm$ 0.3}  \\
				
				Transformer/L1 &  \EP{34.3 $\pm$ 0.2 } & \EP{32.3 $\pm$ 0.3} & \EP{30.9 $\pm$ 0.2} & \EP{29.5 $\pm$ 0.3}  \\
				Transformer/uPL &  \EP{35.7 $\pm$ 0.2} & \EP{34.3 $\pm$ 0.3} & \EP{33.2 $\pm$ 0.4} & \EP{30.0 $\pm$ 0.3}  \\
				\hline
				\hline 
				&\multicolumn{4}{c||}{PSNR - MRA dataset}\\
				\hline 
				DnCNN/L1&  \EP{0.86 $\pm$ 0.01} & \EP{0.81 $\pm$ 0.01} & \EP{31.4 $\pm$ 0.41} & \EP{0.81 $\pm$ 0.01}  \\
				DnCNN/uPL&  \EP{0.92 $\pm$ 0.01} & \EP{0.87 $\pm$ 0.01} & \EP{0.87 $\pm$ 0.01} & \EP{0.84 $\pm$ 0.02}  \\
				ResNet/L1 &  \EP{47.1 $\pm$ 1.6} & \EP{40.1 $\pm$ 0.6} & \EP{39.9 $\pm$ 0.6} & \EP{41.2 $\pm$ 0.6}  \\
				ResNet/uPL &  \EP{44.2 $\pm$ 0.6} & \EP{43.1 $\pm$ 1.2} & \EP{42.6 $\pm$ 0.8} & \EP{40.0 $\pm$ 0.5} \\
				
				Transformer/L1 &  \EP{0.97 $\pm$ 0.03} &\EP{0.91 $\pm$ 0.02} & \EP{0.85 $\pm$ 0.02} & \EP{0.68 $\pm$ 0.03}  \\
				Transformer/uPL & \EP{ 0.98 $\pm$ 0.02} & \EP{0.93 $\pm$ 0.01} & \EP{0.83 $\pm$ 0.01} & \EP{0.78 $\pm$ 0.02}  \\
				
				\hline
				\hline 
				
				\hline
		\end{tabular}}
		\caption{\EP{PSNR values for both datasets}, network structures, and noise levels. }
		\label{table:PSNRnet}
	\end{table}
	
	\renewcommand{\thetable}{S7 Table}
	\begin{table} [tbh!]
		\centering
		\resizebox{0.98\textwidth}{!}
		{
			\begin{tabular}{ || c | c c c  c  || }
				\hline
				&\multicolumn{4}{ c ||}{MSE - MR root dataset}\\ \hline
				
				Network/Loss & 1 \% noise  &   5 \% noise & 10 \% noise & 20 \% noise\\ 
				\hline \hline
				DnCNN/L1&  \EP{5.78e-6 $\pm$ 1e-7} & \EP{6.21e-7 $\pm$ 5e-7} & \EP{7.79e-6 $\pm$ 8e-6} & \EP{7.80e-6 $\pm$ 3e-7}  \\
				DnCNN/uPL&  \EP{5.35e-6 $\pm$ 6e-7} & \EP{5.99e-6 $\pm$ 5e-7} & \EP{7.39e-6 $\pm$ 7e-7} & \EP{7.73e-6 $\pm$ 5e-7}  \\
				ResNet/L1 & \EP{5.92e-6 $\pm$ 7e-7} & \EP{6.51e-7 $\pm$ 4e-7} & \EP{7.68e-6 $\pm$ 4e-6} & \EP{8.01e-6 $\pm$ 5e-7}\\
				ResNet/uPL &  \EP{5.45e-6 $\pm$ 4e-7} & \EP{6.14e-7 $\pm$ 2e-7} & \EP{6.98e-7 $\pm$ 8e-7} & \EP{7.63e-6 $\pm$ 3e-6}  \\
				
				Transformer/L1 &  \EP{5.51e-6 $\pm$ 6e-7} & \EP{5.9e-6 $\pm$ 2e-7} & \EP{7.5e-6 $\pm$ 2e-7} & \EP{7.6e-6 $\pm$ 3e-7}  \\
				Transformer/uPL &  \EP{5.51e-6 $\pm$ 5e-7} & \EP{5.64e-6 $\pm$ 3e-7} & \EP{6.9e-6 $\pm$ 8e-7} & \EP{7.3e-6 $\pm$ 3e-7}  \\

				\hline
				\hline 
				&\multicolumn{4}{ c ||}{MSE (roots) - MR root dataset}\\ \hline
				
				\hline
				DnCNN/L1&  \EP{0.012 $\pm$ 5e-3} & \EP{0.034 $\pm$ 8e-3} & \EP{0.031 $\pm$ 4e-3} & \EP{0.036 $\pm$ 3e-3}  \\
				DnCNN/uPL&  \EP{0.010 $\pm$ 6e-3} & \EP{0.032 $\pm$ 2e-3} & \EP{0.032 $\pm$ 6e-3} & \EP{0.032 $\pm$ 7e-3}  \\
				ResNet/L1 &  \EP{0.013 $\pm$ 7e-3} & \EP{0.033$\pm$ 8e-3} & \EP{0.034 $\pm$ 6e-3} & \EP{0.034 $\pm$ 8e-3}  \\
				ResNet/uPL &  \EP{0.0098 $\pm$ 6e-3} & \EP{0.031 $\pm$ 3e-3} & \EP{0.032 $\pm$ 5e-3} & \EP{0.033 $\pm$ 6e-3}  \\
				
				Transformer/L1 &  \EP{0.012 $\pm$ 2e-3 } & \EP{0.035 $\pm$ 9e-3} & \EP{0.036 $\pm$ 6e-3} & \EP{0.036 $\pm$ 8e-3}  \\
				Transformer/uPL &  \EP{0.010 $\pm$ 4e-3} & \EP{0.032 $\pm$ 5e-3} & \EP{0.033 $\pm$ 6e-3} & \EP{0.034 $\pm$ 5e-3}  \\

				\hline
				\hline 
				&\multicolumn{4}{c||}{MSE - MRA dataset}\\
				\hline 
				DnCNN/L1&  \EP{4.1e-3 $\pm$ 6e-4} & \EP{4.7e-3 $\pm$ 4e-4} & \EP{0.034 $\pm$ 0.004} & \EP{5.9e-3 $\pm$ 5e-4}  \\
				DnCNN/uPL&  \EP{3.8e-3 $\pm$ 5e-4} & \EP{4.1e-3 $\pm$ 5e-4} & \EP{4.9e-3 $\pm$ 4e-4} & \EP{5.3e-3 $\pm$ 5e-4}  \\
				ResNet/L1 &  \EP{3.2e-3 $\pm$ 5e-4} & \EP{4.8e-3 $\pm$ 3e-4} & \EP{5.9e-3 $\pm$ 7e-4} & \EP{5.1e-3 $\pm$ 6e-4}  \\
				ResNet/uPL &  \EP{3.9e-3 $\pm$ 4e-4} & \EP{4.5e-3 $\pm$ 2e-4} & \EP{ 4.9e-3 $\pm$ 6e-4} & \EP{5.7e-3 $\pm$ 5e-4} \\
				
				Transformer/L1 &  \EP{3.8e-3 $\pm$ 5e-4} &\EP{4.3e-3 $\pm$ 8e-4} & \EP{5.7e-3 $\pm$ 5e-4} & \EP{6.2e-3 $\pm$ 3e-4}  \\
				Transformer/uPL & \EP{3.1e-3 $\pm$ 4e-4} & \EP{4.1e-3 $\pm$ 5e-4} & \EP{6.1e-3 $\pm$ 7e-4} & \EP{6.2e-3 $\pm$ 8e-4}  \\
				
				\hline
				\hline 
				
				\hline
		\end{tabular}}
		\caption{\EP{MSE values only calculated on the roots part and MSE values for both} datasets for all network architectures, and noise levels included in this study. }
		\label{table:MSEnet}
	\end{table}

\renewcommand{\thetable}{S8 Table}
\begin{table} [tbh!]
			\centering
			{
				\begin{tabular}{ || c | c c c  c  || }
					\hline
					&\multicolumn{4}{ c ||}{SSIM - Image Part MR root}\\ \hline
					
					Network/Loss & 1 \% noise  &   5 \% noise & 10 \% noise & 20 \% noise\\ 
					\hline \hline
					DnCNN/L1&  \EP{0.80 $\pm$ 0.01  } & \EP{0.77 $\pm$ 0.01} & \EP{0.73 $\pm$ 0.01 } & \EP{0.74 $\pm$ 0.01}  \\
					DnCNN/uPL&  \EP{0.85 $\pm$ 0.01  } & \EP{0.84 $\pm$ 0.01} & \EP{0.77 $\pm$ 0.01 } & \EP{0.76 $\pm$ 0.01} \\
					ResNet/L1 &  \EP{0.81 $\pm$ 0.02} & \EP{0.77 $\pm$ 0.01} & \EP{0.75 $\pm$ 0.01} & \EP{0.74 $\pm$ 0.01}  \\
					ResNet/uPL &  \EP{0.85 $\pm$ 0.01} & \EP{0.83 $\pm$ 0.01} & \EP{0.78 $\pm$ 0.01} & \EP{0.78 $\pm$ 0.01}  \\
					
					Transformer/L1 &  \EP{0.77 $\pm$ 0.02 } & \EP{0.75 $\pm$ 0.01} & \EP{0.67 $\pm$ 0.02} & \EP{0.46 $\pm$ 0.03}  \\
					Transformer/uPL &  \EP{0.81 $\pm$ 0.01} & \EP{0.79 $\pm$ 0.01} & \EP{0.77 $\pm$ 0.01} & \EP{0.66 $\pm$ 0.02} \\

					\hline
					\hline 
					&\multicolumn{4}{c||}{SSIM - Image Part MRA}\\
					\hline 
					DnCNN/L1&  \EP{0.95 $\pm$ 0.01} & \EP{0.87 $\pm$ 0.01} & \EP{0.83 $\pm$ 0.01} & \EP{0.83 $\pm$ 0.02}  \\
					DnCNN/uPL&  \EP{0.97 $\pm$ 0.01} & \EP{0.92 $\pm$ 0.01} & \EP{0.87 $\pm$ 0.01} & \EP{0.86 $\pm$ 0.01} \\
					ResNet/L1 &  \EP{0.98 $\pm$ 0.01} & \EP{0.92 $\pm$ 0.01} & \EP{0.87 $\pm$ 0.01} & \EP{0.84 $\pm$ 0.01}  \\
					ResNet/uPL &  \EP{0.98 $\pm$ 0.01} & \EP{0.95 $\pm$ 0.01} & \EP{0.89 $\pm$ 0.01} & \EP{0.87 $\pm$ 0.01} \\
					
					Transformer/L1 &  \EP{0.90 $\pm$ 0.01} &\EP{0.86 $\pm$ 0.01} & \EP{0.84 $\pm$ 0.01} & \EP{0.78 $\pm$ 0.01}  \\
					Transformer/uPL & \EP{0.95 $\pm$ 0.01} & \EP{0.89 $\pm$ 0.01} & \EP{0.83 $\pm$ 0.01} & \EP{0.83 $\pm$ 0.01}  \\
					
					\hline
					\hline 
					
					\hline
			\end{tabular}}
			\caption{\EP{ SSIM values for both datasets calculated on the center of the image. }}
			\label{table:PartSSIM}
		\end{table}

\renewcommand{\thetable}{S9 Table}
\begin{table} [tbh!]
			\centering
			{
				\begin{tabular}{ || c | c c c  c  || }
					\hline
					&\multicolumn{4}{ c ||}{PSNR - Image Part MR root}\\ \hline
					
					Network/Loss & 1 \% noise  &   5 \% noise & 10 \% noise & 20 \% noise\\ 
					\hline \hline
					DnCNN/L1&  \EP{35.6 $\pm$ 0.2  } & \EP{35.5 $\pm$ 0.4} & \EP{34.2 $\pm$ 0.3 } & \EP{34.0 $\pm$ 0.2}  \\
					DnCNN/uPL&  \EP{35.7 $\pm$ 0.3 } & \EP{34.8 $\pm$ 0.2} & \EP{35.5 $\pm$ 0.4 } & \EP{34.3 $\pm$ 0.3} \\
					ResNet/L1 &   \EP{35.7 $\pm$ 0.2  } & \EP{34.3 $\pm$ 0.2} & \EP{35.3 $\pm$ 0.2 } & \EP{33.3 $\pm$ 0.3}  \\
					ResNet/uPL &  \EP{36.2 $\pm$ 0.2 } & \EP{35.9 $\pm$ 0.3} & \EP{35.5 $\pm$ 0.3 } & \EP{34.3 $\pm$ 0.3} \\
					
					Transformer/L1 &  \EP{35.4 $\pm$ 0.2 } & \EP{31.3 $\pm$ 0.3} & \EP{29.8 $\pm$ 0.4} & \EP{27.7 $\pm$ 0.4}  \\
					Transformer/uPL &  \EP{35.8 $\pm$ 0.2} & \EP{34.3 $\pm$ 0.4} & \EP{31.3 $\pm$ 0.3} & \EP{29.8 $\pm$ 0.3}  \\

					\hline
					\hline 
					&\multicolumn{4}{c||}{PSNR - Image Part MRA}\\
					\hline 
					DnCNN/L1&  \EP{37.0 $\pm$ 0.2} & \EP{32.8 $\pm$ 0.3} & \EP{30.2 $\pm$ 0.4} & \EP{28.0 $\pm$ 0.4}  \\
					DnCNN/uPL&  \EP{37.4 $\pm$ 0.3} & \EP{33.4 $\pm$ 0.2} & \EP{33.9 $\pm$ 0.3} & \EP{30.1 $\pm$ 0.2} \\
					ResNet/L1 &  \EP{37.3 $\pm$ 0.2} & \EP{37.0 $\pm$ 0.2} & \EP{34.2 $\pm$ 0.3} & \EP{28.2 $\pm$ 0.4}  \\
					ResNet/uPL &  \EP{39.3 $\pm$ 0.2} & \EP{38.6$\pm$ 0.3} & \EP{37.8 $\pm$ 0.3} & \EP{32.1 $\pm$ 0.3} \\
					
					Transformer/L1 &  \EP{39.1 $\pm$ 0.4} &\EP{38.0 $\pm$ 0.3} & \EP{32.2 $\pm$ 0.3} & \EP{26.2 $\pm$ 0.3}  \\
					Transformer/uPL & \EP{39.0 $\pm$ 0.3} & \EP{37.5 $\pm$ 0.2} & \EP{34.3 $\pm$ 0.2} & \EP{32.2 $\pm$ 0.4}  \\
					
					\hline
					\hline 
					
					\hline
			\end{tabular}}
			\caption{\EP{ PSNR values for both datasets calculated on the center of the image. }}
			\label{table:PartPSNR}
		\end{table}

\renewcommand{\thetable}{S10 Table}
\begin{table} [tbh!]
			\centering
			
				\begin{tabular}{ || c | c c c  c  || }
					\hline
					&\multicolumn{4}{ c ||}{MSE (roots) - Image Part MR root}\\ \hline
					
					Network/Loss & 1 \% noise  &   5 \% noise & 10 \% noise & 20 \% noise\\ 
					\hline \hline
					DnCNN/L1&  \EP{0.008 $\pm$ 7e-4  } & \EP{0.012 $\pm$ 1e-3} & \EP{0.012 $\pm$ 1e-3 } & \EP{0.019 $\pm$ 1e-3}  \\
					DnCNN/uPL&  \EP{0.006 $\pm$ 8e-4 } & \EP{0.009$\pm$ 9e-4} & \EP{0.015 $\pm$ 7e-4 } & \EP{0.017 $\pm$ 1e-4} \\
					ResNet/L1 &   \EP{0.007 $\pm$ 6e-4  } & \EP{0.009 $\pm$ 8e-4} & \EP{0.010 $\pm$ 7e-4 } & \EP{0.011 $\pm$ 7e-4}  \\
					ResNet/uPL &  \EP{0.006 $\pm$ 8e-4 } & \EP{0.007 $\pm$ 1e-3} & \EP{0.008 $\pm$ 8e-4 } & \EP{0.008 $\pm$ 8e-4} \\
					Transformer/L1 &  \EP{0.010 $\pm$ 9e-4 } & \EP{0.018 $\pm$ 2e-3} & \EP{0.019 $\pm$ 1e-3} & \EP{0.021 $\pm$ 2e-3}  \\
					Transformer/uPL &  \EP{0.011 $\pm$ 8e-4} & \EP{0.012 $\pm$ 1e-3} & \EP{0.014 $\pm$ 2e-3} & \EP{0.019 $\pm$ 2e-3}  \\
					\hline
					\hline 
					&\multicolumn{4}{c||}{MSE - Image Part MRA}\\
					\hline 
					DnCNN/L1&  \EP{0.0017 $\pm$ 6e-4 } & \EP{0.0025 $\pm$ 7e-4 } & \EP{0.0031 $\pm$ 6e-4} & \EP{0.0033 $\pm$ 8e-4}  \\
					DnCNN/uPL&  \EP{0.0016 $\pm$ 5e-4} & \EP{0.0019 $\pm$ 8e-4} & \EP{0.0023 $\pm$ 7e-4 } & \EP{0.0025 $\pm$ 9e-4 }  \\
					ResNet/L1 &  \EP{0.0013 $\pm$ 7e-4 } & \EP{0.0021 $\pm$ 6e-4 } & \EP{0.0024 $\pm$ 7e-4 } & \EP{0.0025 $\pm$ 7e-4 }  \\
					ResNet/uPL &  \EP{0.0011 $\pm$ 5e-4 } & \EP{0.0017 $\pm$ 8e-4 } & \EP{ 0.0020 $\pm$ 5e-4} & \EP{0.0021 $\pm$ 8e-4 } \\
					
					Transformer/L1 &  \EP{0.0015 $\pm$ 4e-4 } &\EP{0.0019 $\pm$ 5e-4 } & \EP{0.0039 $\pm$ 4e-4 } & \EP{0.0043 $\pm$ 6e-4 }  \\
					Transformer/uPL & \EP{ 0.0016 $\pm$ 6e-4} & \EP{0.0017  $\pm$ 6e-4} & \EP{0.0021 $\pm$ 9e-4 } & \EP{0.0022 $\pm$ 9e-4 }  \\
					\hline
					\hline 
					
					\hline
			\end{tabular}
			\caption{\EP{ MSE values for both datasets calculated on the center of the image. }}
			\label{table:PartMSE}
		\end{table}

\end{document}